# Word Frequency Counting Based on Serverless MapReduce

Hanzhe Li[1][a], Bingchen Lin[2][b], Mengyuan Xu[3][c]

[1] College of Artificial Intelligence, Xi'an Jiaotong University, Xianning West Road, Xi'an, Shaanxi Province, China
[2] College of Artificial Intelligence, Chongqing University of Education, Chongjiao Road, Chongqing, China
[3] College of Computer and Information Engineering, Qilu Institute of Technology, Jingshi East Road, Jinan City, Shandong Province, China
[1] lhz2023@stu.xjtu.edu.cn, [2] 2311401218@stu.cque.edu.cn, [3] 20202270@stu.hebmu.edu.cn

Keywords: Serverless Computing; MapReduce; Word Frequency Counting; Cloud Computing

Abstract: With the increasing demand for high-performance and high-efficiency computing, cloud computing, especially serverless computing, has gradually become a research hotspot in recent years, attracting numerous research attention. Meanwhile, MapReduce, which is a popular big data processing model in the industry, has been widely applied in various fields. Inspired by the serverless framework of Function as a Service and the high concurrency and robustness of MapReduce programming model, this paper focus on combining them to reduce the time span and increase the efficiency when executing the word frequency counting task. In this case, the paper use a MapReduce programming model based on a serverless computing platform to figure out the most optimized number of Map functions and Reduce functions for a particular task. For the same amount of workload, extensive experiments show both execution time reduces and the overall efficiency of the program improves at different rates as the number of map functions and reduce functions increases. This paper suppose the discovery of the most optimized number of map and reduce functions can help cooperations and programmers figure out the most optimized solutions.

## 1 INTRODUCTION

In order to meet the growing demand for computing resources and high-end chipsets in real-world applications (McGrath et al., 2017; Baldini et al., 2017), cloud computing technology has attracted increasing research interest in recent years, forming various classic cloud service models such as Infrastructure as a Service (IaaS), Platform as a Service (PaaS), and Software as a Service (SaaS). However, these models mentioned above rely on high levels of professional knowledge, which are costly and cannot achieve a balance between management, expansion, and cost-effectiveness indicators. In order to alleviate the above problems, serverless computing has emerged, aiming to reduce the burden of server management and save cloud service costs (Vincent et al., 2019).

The basic unit of serverless computation is a function. When receiving a user request, the serverless platform calls the relevant functions on the platform based on the parameters in the request, such as the URL of the function. This service model is commonly referred to as Function as a Service (FaaS), which is usually paired with the Backend as a Service (BaaS). Compared with traditional centralized monolithic applications, FaaS services are composed of independent functions explicitly arranged, which can intuitively represent the business logic control and data flow of the application. Additionally, serverless computing is much more economical and cost-friendly as users no longer need to pay for extra idled computing resources, the maintenance of used resources as well as the security of the used resources.

[a] https://orcid.org/0009-0002-8999-7996

[b] https://orcid.org/0009-0001-8866-7752

[c] https://orcid.org/0009-0005-0411-3656

Serverless computing enables users to focus more on the logic of their programs. As for the maintenance of the backend servers, it is all up to the service provider. (McGrath et al., 2017; Jeffrey et al., 2004) Serverless computing features more scalability and elasticity than traditional local computing servers, since the dynamic allocation of computing resources makes it possible for users to handle sudden surge in workloads and data processing demands. Currently, there are many serverless computing platforms that provides state-of-the-art cloud computing services, such as AWS Lambda, Google Cloud, Microsoft Azure, Alibaba Cloud etc.

MapReduce is currently the most popular model for processing massive amounts of data, which mainly includes four stages: Map, Partition, Shuffle, and Reduce. MapReduce is widely used for parallel processing across distributed systems and generating large-scale datasets. First, it is user-friendly, even for beginners, as it conceals the specific intricacies involving parallelization, fault-tolerance, optimizing locality, and balancing workloads. Second, many complex problems in the real world are highly expressible in the MapReduce programming model, such as word counting, word frequency analysis etc. (Baldini et al., 2017) However, MapReduce is often constrained by the data transmission method. Specifically, due to the need for the mapper to be completed as soon as possible, there may be a risk of timeout for the mapper while the reducer is still working. Therefore, it is not feasible to directly transfer data between mappers and reducers. In this context, combining serverless and MapReduce frameworks shows promising application prospects.

Inspired by these two cutting-edge and matured technologies, this paper focus on combining them to reduce the time span and increase the efficiency when executing the word frequency counting task. This paper uses a MapReduce programming model based on a serverless computing platform to figure out the most optimized number of Map functions and Reduce functions. Though it seemed obvious that the more map and reduce functions are implemented, the higher the overall efficiency the program may achieve. This paper's goal, however, is to figure out the trend at which the overall efficiency is increasing. The results indicate that, when executing the same amount of workloads, as the number of map functions and reduce functions increases, both execution time reduces and the overall efficiency of the program improves but at different rates. This paper hopes to find out the most optimized number of map and reduce functions so as to help cooperations and programmers figure out the most optimized solutions when implementing the MapReduce programming model on their tasks and workflows.

Focusing on above aspects, this paper starts with a brief overview of the basic principles of the MapReduce programming model, the operating rules of serverless computing platforms as well as services and the overall framework of the experiment (Section 2). Then, the paper discusses relevant methodologies as well as evaluations and presents the result of the experiment conducted by giving in-depth evaluations and conclusions based on existing research data and results (Section 3). Lastly, the paper discusses current drawbacks of the experiment framework used in this paper, analyses the strengths and weaknesses of the results and envisions possible solutions and new research areas based on current experiments (Section 4). This paper also summarizes in Section 4.

## 2 METHOD

### 2.1 Revisiting MapReduce and Serverless

In this section, the paper presents a brief overview of the basic principles of MapReduce programming model as well as the operating rules of the serverless computing platform.

**MapReduce**. The overall MapReduce programming model mainly consists of two functions, two phases as well as three categories of files. In terms of three categories of files, there are input files, intermediate files as well as the output files. The input file contains data that needs to be processed. The intermediate files contain important data that are needed during the MapReduce executing process and the output files hold the final result of the program. In terms of the two functions and two phases, there is the Map function, which relates to the Map phase, and the Reduce function, which relates to the Reduce phase. The Map function is responsible for reading data from the input files and process these data into key-value pairs, which are later stored in intermediate files. These intermediate files forward these key-value pairs to the Reduce function, where these key-value pairs are sorted, partitioned and processed into final results and are written into the output files, which later are available and accessible to the user (Jeffrey et al., 2004).

**Serverless**. The operating rules of serverless computing platform consists of four main stages, which are: Event Trigger, Function Execution, Function Processing and Response Return. In the Event Trigger stage, there is a local client, which runs locally on the user's device. The client triggers an event, such as an HTTP request, file upload or a message queue. Then, the trigger passes the event to the function on the cloud, entering Function Execution stage. Once the function on the cloud is triggered, the serverless computing platform will dynamically allocate and scale the computing resources to start executing the function. In the Function Processing stage, codes in functions are executed and outputs are generated. During this process, the function on the platform will be authorized to access and manipulate the storage, databases and other related services requested by the user in advance. Lastly, in the Response Return stage, the function returns the output to the local client of the user. The response can be anything, such as response data, state updates and notifications in various forms etc. Often, the results generated by the functions are stored in the storage services provided by the serverless computing platform.

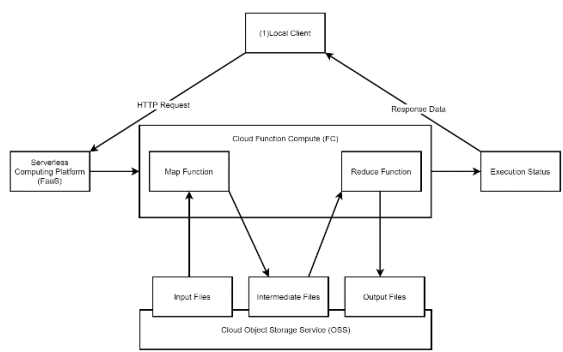

Figure 1: The framework of proposed method.

## 2.2 Overall Framework

The main goal of this paper is word frequency analysis using MapReduce based on serverless computing. To perfectly combine MapReduce programming model and serverless computing and word frequency analysis altogether, this paper implemented the following methods and made miniscule changes to the MapReduce programming model.

The entire experiment is firstly conducted with controlled variables method. This paper manages to analyse the same set of word documents, which are in the text document format, but use MapReduce frameworks in different parameters. The parameters are different in areas such as the number of MapReduce functions, the configuration of CPUs and RAMs on the serverless network etc. Therefore, during the entire process of the experiment, performances can be analysed via the changes applied to these parameters.

Secondly, here is the devices used in the entire experiment process. As is shown in Figure 1, the serverless MapReduce framework of this paper contains a local python client, which is deployed in PyCharm. This client is responsible for calling functions deployed on the serverless platform and receiving completion signals once MapReduce functions are executed successfully. The serverless computing platform used during the experiment is Alibaba Cloud Platform. The services this paper uses in particular is the Alibaba Cloud Function Compute (FC), where the team deploys MapReduce functions, and Alibaba Cloud Object Storage Service (OSS), where the team stores the files related to this experiment temporarily so that any process that requires reading and writing files stays on the serverless platform, ensuring that data transfer speeds between local and cloud does not affect the execution time significantly.

Lastly, the to-be-analysed files this paper uses are of the same quality. Each file is roughly about 1,000,000. It is critical to keep the word count of these files roughly the same, as different workloads can also contribute to the performance difference of each test.

## 2.3 MapReduce Functions

In terms of the miniscule changes to the MapReduce functions, this paper customized how Map functions read the data. Each Map function in this experiment contains a set of parameters, which are "file ids", "number of files" and "index". These parameters help the Map functions read the correct group of files stored in the Alibaba Cloud OSS so as to make sure that each file is only processed once throughout the execution.

## 3 EXPERIMENT

### 3.1 Experiment Settings

In the field of modern technology, the improvement of computing speed has always been a focus of attention for researchers and technical engineers (Zhenyu et al., 2023). In order to achieve more efficient computing, this paper adopted a new strategy in this experiment (Jeffrey et al., 2004), which is to use multi-threaded technology to replace single threading, in order to optimize computing speed. Multi-threading technology, in simple terms, means executing multiple tasks simultaneously to complete more work at the same time. Compared to single threading, multi-threading can complete more tasks in a relatively short period of time, thereby improving overall computing speed. This technology has achieved significant results in the field of computer science, especially in processing large complex mathematical models, large-scale data analysis, and real-time communication, with significant advantages.

Since this paper aim to investigate the serverless MapReduce based on the application of multi-threading technology to enhancing computing speed, the team first analyzed existing single-threaded programs and identified the bottleneck parts that require optimization in subsequent multi-threaded designs. Subsequently, the team devised corresponding multi-threaded algorithms and conducted detailed analysis and testing on them. Throughout the experiment, the team continuously adjusted and refined the multi-threading strategy to achieve the most significant improvement in computing speed. There is an encoding file named client written locally on the computer. The team found FC in the console and created a function in its service. In the Function Services, the team have written down the map function and the reduce function.

In the experiment, there were 50 files that were used in the experiment. Each of the files contained roughly about 1,000,000 words. The files were initially uploaded to Alibaba Cloud Object Storage Service (OSS) using a local string of code and network protocols stored in OSS. Then the team create the Mapper functions and Reducer functions in advance in Alibaba Cloud Function Compute (FC). Each function on the platform is deployed on vCPU 0.35 with 512MB of RAM configured. Later, the team enable pre-prepared client code locally, allowing 50 files stored in OSS to be called into Alibaba Cloud Function Compute (FC) so as to start the program running.

Previous works (J Jiang et al., 2021; Prasoon et al., 2024) have shown that it is plausible to evaluate the MapReduce programming model and serverless computing performance based on their execution timespan and memory usage. For one thing, execution time is the direct reflection of the performance of the program. For another, memory usage implies the resource management and allocation during the execution, enabling the team to observe the results in a clearer way. Furthermore, the team are able to optimize the workloads assigned to each function and enhance the algorithms simultaneously, therefore improving the methods throughout the experiment process (Rodrigo et al., 2024; Q Liu et al., 2024).

Table 1: Model performance comparison under different numbers of MapReduce functions (with 50 files)

| Func Num | Average Execution Time /ms | | Average RAM Usage /MB | |
|---|---|---|---|---|
| | Mapper | Reducer | Mapper | Reducer |
| 1 | 40816.58 | 51624.64 | 1604.26 | 1021.02 |
| 2 | 7716.69 | 15133.26 | 821.54 | 533.82 |
| 5 | 2455.69 | 4269.94 | 351.06 | 246.82 |
| 10 | 1464.97 | 2198.27 | 194.01 | 139.54 |

## 3.2 RAM Usage for Different Numbers of MapReduce Functions

The team first quantitatively compared the impact of different MapReduce functions on RAM usage, whose results are shown in Table 1. In the first case of the experiment, the team utilized only one MapReduce function. The average execution time of the Mapper function was 40816.58ms, the average execution time of the Reducer function was 51624.64ms, and the RAM utilized by the Mapper and Reducer amounted to 1021.02 MB and 1604.26 MB, respectively. In the following case of the experiment, two sets of MapReduce functions are deployed. The average execution time of the mapper function is 7716.69ms. The average execution time of the Reducer function is 15133.26ms. The RAM used by the Mapper and Reduce is 533.82MB and 821.54MB. In the third case of the experiment, the team use five MapReduce functions. The average execution time of the Mapper function is 2455.694ms, and the average execution time of the Reducer function reached 4269.94ms. The RAMs used by Mapper and Reducer are 246.824MB and 351.066MB. In the last case of the experiment, the team use 10 MapReduce functions. The average execution time of the Mapper function is 1464.974ms and the average execution time of the Reducer function reached 2198.27ms. The RAMs used by Mapper and Reducer are 139.545MB and194.016MB.

## 3.3 Time Cost for Different Numbers of MapReduce Functions

As shown in Figure 2, the results indicate that as the number of MapReduce functions increases, the average execution time gradually decreases. This indicates that in the process of big data processing, increasing the number of MapReduce functions reasonably can effectively improve the efficiency of data processing and reduce execution time (J Cai et al., 2023). However, these results also indicate that improving the number of MapReduce functions aimlessly is not an effective way, since the rates at which the execution time is decreasing are dropping. So, the team come to a brief conclusion that when configuring the number of MapReduce functions, it is best to suit the workload and the existing resources, as this way can generate the most ideal result possible without consuming too much resources or being too costly.

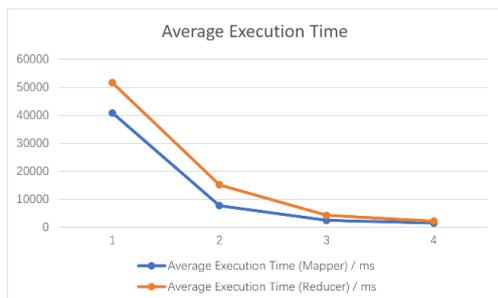

Figure 2: Average Execution Time of MapReduce Functions.

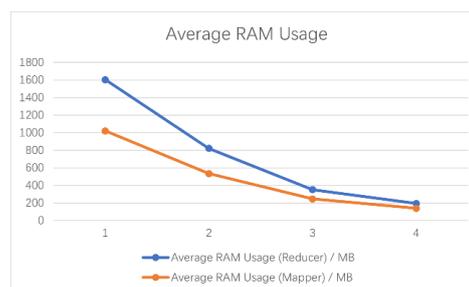

Figure 3: Average RAM Usage of Functions.

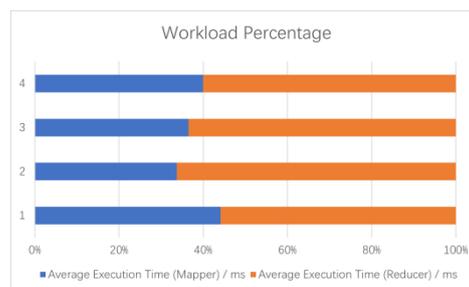

Figure 4: Workload Percentage of MapReduce Functions.

### 3.4 Comparison for Memory Usage and Average Usage Time

After completing all the experimental work, the team focused on the memory usage during the runtime of the MapReduce function. As the number of MapReduce functions increases in Figure 3, the average memory usage also shows a decreasing trend. This may be because as the number of functions increases, the system can execute tasks in parallel on more cores, thereby reducing the memory footprint of individual tasks. In addition, by optimizing the writing and execution strategies of the MapReduce function, the team can further reduce memory usage and improve system resource utilization.

The team analyses the average usage time ratio of Mapper and Reducer in each experiment. Through comparison, the team found that it cannot be simply assumed that as the number of MapReduce functions increases, the time consumed by Mapper processing data will become longer. During the experiment, the team use different numbers of MapReduce functions to conduct detailed timing analysis for each experiment. The results showed that there were certain differences in the proportion of usage time between Mapper and Reducer in different experiments. This indicates that an increase in the number of MapReduce functions does not necessarily lead to a decrease in Mapper processing time.

### 3.5 Impact Analysis for the MapReduce Implementation

As is Figure 4 shown above, the team also explored the impact of the internal implementation of the MapReduce function on the time ratio of Mapper and Reduce usage. By comparing the code implementations in different experiments, the team found that the optimization of the internal implementation of the MapReduce function may change the usage time ratio of Mapper and Reduce. This means that when increasing the number of MapReduce functions, optimizing the internal implementation can effectively reduce the processing time of the Mapper, thereby improving overall computational efficiency. To this end, in order to improve overall computing performance, the team need to pay attention to data size, internal implementation of MapReduce function, and other influencing factors, in order to achieve more efficient distributed computing in practical applications.

## 4 CONCLUSIONS AND FUTURE WORKS

This paper has introduced multi-threaded MapReduce framework based on serverless computing platforms in order to boost overall efficiency of the program as well as minimizing local server maintenance thanks to the user-friendly serverless computing platform. When it comes to analysing word frequency with MapReduce based on serverless platform, the team first identified that as the number of MapReduce functions, also referred to as thread, increases, the speed at which the program is executing increases simultaneously. However, there is a peak in the rate at which the speed is increasing, meaning that increasing the number of MapReduce functions aimlessly is likely to result in a waste of computing resources or lead to lower efficiency in utilizing the serverless computing resources.

To this aim, this paper conducted a series of experiment of serverless MapReduce in terms of MapReduce function numbers and assess the results based on the average execution time and average memory usage during the execution. The team speculated that the rate at which the execution time is dropping experiences a major drop and then starts to slow down. Therefore, the team come to a conclusion that increasing the number of MapReduce functions aimlessly does not always contribute to the efficiency of the program, and that for different tasks, the number of MapReduce functions should be calculate respectively and carefully so as to utilize the computing resources to its full potential (Prasoon et al., 2024).

The team also notice that there are some limitations when extending this work to real-life applications, which mainly comes from the ideal setting that each word file has the approximately the same workload. Besides, the word frequency analysis task the team perform is not universally reliable, as it is a low demanding task in terms of computing resources. Therefore, furthermore types of tasks are required to complete the research. The team's future work includes how to dynamically allocate MapReduce functions to different workloads so that for each word file, a sufficient number of MapReduce function is implemented in order to achieve a better efficiency when executing a task that is not evenly distributed among files.

## AUTHORS CONTRIBUTION

All the authors contributed equally and their names were listed in alphabetical order.